# NUMERICAL SEARCH FOR UNIVERSAL ENTANGLERS IN $C^3 \otimes C^4$ and $C^4 \otimes C^4$

F. V. Mendes,   R. V. Ramos

fernandovm@gmail.com   rubens.viana@pq.cnpq.br

[1]Lab. of Quantum Information Technology, Department of Teleinformatic Engineering – Federal University of Ceara - DETI/UFC, C.P. 6007 – Campus do Pici - 60455-970 Fortaleza-Ce, Brazil.

A universal entangler is quantum gate able to transform any disentangled state in an entangled state. Although universal entanglers are abundant in arbitrary high dimensional spaces, to verify if a quantum gate is a universal entangler is a hard task since it is not known which property of the unitary matrix is responsible for such behavior. In this direction, the present work shows the results of an algorithm based on differential evolution that tests universal entanglers in $C^3 \otimes C^4$ and $C^4 \otimes C^4$. We present two good candidates for each cited space and we show that a candidate found in the literature is not a universal entangler.

## 1. Introduction

Quantum entanglement plays a fundamental role in many interesting tasks such as quantum teleportation, quantum dense coding and other quantum cryptography protocols [1,2]. In this context, a relevant attention has been devoted to the study of entanglers [3-6], aiming to understand their properties, construction and applications. A particularly interesting class of entanglers is the universal entangler, a quantum gate able to transform any disentangled state (belonging to the appropriated Hilbert space) in an entangled state. Although universal entanglers are known to be abundant, it is not an easy task to affirm that a given quantum gate is a universal entangler since, in general, it is not known which property of the unitary matrix is responsible for such behavior. Furthermore, due to the dimensions of the spaces considered, a brute-force checking is not viable. An alternative is to use some heuristic in order to implement an intelligent search for universal entanglers. Even known that such heuristics are not able to confirm that a given gate is in fact a universal entangler, the use of heuristics has two interesting advantages: 1) It can find good candidates for universal entanglers. 2) It can be useful to discard pseudo-good candidates. In this direction, the present work shows the results obtained by an algorithm based on differential evolution that tests universal entanglers. We provide two good candidates for $C^3 \otimes C^4$ and two good candidates for $C^4 \otimes C^4$. Moreover, we also show that a candidate for a universal entangler presented in the literature is not a universal entangler.

## 2. Separability

In the core of the proposed algorithm are the separability test of quantum states and the entanglement measure of pure bipartite states. There are good bipartite entangled measures like partial von Neumann entropy [7], concurrence [8], positive partial transpose (PPT) [9], negativity

[10] among others. In this work only the entanglement measure based on the von Neumann entropy will be used. Given an entangled pure bipartite state $|\varphi\rangle_{AB}$, its entanglement is given by

$$E_{VN}\left(|\varphi\rangle_{AB}\langle\varphi|\right) = S_{VN}\left(Tr_A\left(|\varphi\rangle_{AB}\langle\varphi|\right)\right) = S_{VN}\left(Tr_B\left(|\varphi\rangle_{AB}\langle\varphi|\right)\right) \tag{1}$$

$$S_{VN}(\rho) = -Tr[\rho \ln(\rho)]. \tag{2}$$

In what concerns the separability, one says that a given normal matrix $N \in C^{m \cdot n}$ is separable if it can be described as the Kronecker product of two other normal matrices, $N = N_A \otimes N_B$, where $N_A \in C^m$ and $N_B \in C^n$. However, since the Kronecker product is a bilinear form, one can define the map $\gamma: C^m \times C^n \to C$ as

$$\gamma(\vec{p},\vec{q},\vec{r},\vec{s}) = (\vec{p} \otimes \vec{r})^\dagger N (\vec{q} \otimes \vec{s}). \tag{3}$$

In (3) the column vectors $\vec{p}, \vec{q} \in C^m$ while the column vectors $\vec{r}, \vec{s} \in C^n$. Thus, a necessary and sufficient condition to have $N$ decomposable in the Kronecker product $N_A \otimes N_B$ is

$$\gamma(\vec{p}_1,\vec{q}_1,\vec{r}_1,\vec{s}_1) \cdot \gamma(\vec{p}_2,\vec{q}_2,\vec{r}_2,\vec{s}_2) = \gamma(\vec{p}_1,\vec{q}_1,\vec{r}_2,\vec{s}_2) \cdot \gamma(\vec{p}_2,\vec{q}_2,\vec{r}_1,\vec{s}_1) \tag{4}$$

to be true for all vectors $\vec{p}_i, \vec{q}_i$ in a basis of $C^m$ and $\vec{r}_j, \vec{s}_j$ in a basis of $C^n$. This approach was used in [11], for unitary matrices, in order to determine the conditions of separability preservation under conjugation of two-qubit quantum gates. When (3)-(4) is used to test the separability of the Hermitean matrix corresponding to a general two-qubit state, $\alpha_1|00\rangle+\alpha_2|01\rangle+\alpha_3|10\rangle+\alpha_4|11\rangle$, the known result $|\alpha_1\alpha_4-\alpha_2\alpha_3|=0$ is obtained. In general, the usage of (3)-(4) to check the biseparability of quantum states results in a set of polynomial equations in the states' coefficients that have to identically null. If anyone of them is not equal to zero, then the state is not biseparable. Fort example, it was affirmed in [12] that the three-qutrit state

$$|\kappa\rangle = \frac{1}{\sqrt{6}}\left(|000\rangle - |011\rangle - |112\rangle + |120\rangle - |202\rangle + |221\rangle\right)_{ABC} \tag{5}$$

is biseparable. However, using (3)-(4), the conditions of biseparability of a general three-qutrit state $\alpha_1|000\rangle+\alpha_2|001\rangle+\alpha_3|002\rangle+\ldots+\alpha_{27}|222\rangle$ include the following equations

$$\alpha_1^2 \cdot \alpha_1^* \cdot \alpha_{15}^* - \alpha_1^2 \cdot \alpha_6^* \cdot \alpha_{10}^* = 0 \quad \text{for A\_BC} \tag{6}$$

$$\alpha_1^2 \cdot \alpha_1^* \cdot \alpha_{11}^* - \alpha_1^2 \cdot \alpha_2^* \cdot \alpha_{10}^* = 0 \quad \text{for B\_AC} \tag{7}$$

$$\alpha_1^2 \cdot \alpha_1^* \cdot \alpha_{13}^* - \alpha_1^2 \cdot \alpha_4^* \cdot \alpha_{10}^* = 0 \quad \text{for AB\_C.} \tag{8}$$

Using $\alpha = 1/6^{1/2}$ in (6)-(8), one gets for the three cases the value -1/36, hence, the state in (5) is not biseparable.

## 3. Universal Entanglers

Any quantum gate able to generate entanglement between two separable states is considered an entangler. The concept of entangling power was developed in [13] aiming to characterize how good entangler is a particular quantum gate. In [5] it was established a connection between entangling power and the invariants of two-qubit gates. In [14] it was introduced the perfect entangler concept: perfect entanglers are entangler gates able to generate maximally entangled states from separable states. Another entangler class was defined in [15] aiming to respond the question: Does there exist any perfect entangler which can maximally entangle a full separable basis? This particular class has received the name of the especial perfect entanglers. A still more general definition about entanglers was introduced in [16], named universal entangler, in response to the question: Does there exist an entangler able to entangle any separable state? Formally, a universal entangler is a quantum gate $U$ such that $E_{VN}(U(|\varphi\rangle_m|\psi\rangle_n)) > 0$, where $|\varphi\rangle_m$ and $|\psi\rangle_n$ are arbitrary one qudit states of dimensions $m$ and $n$, respectively. In [16] it was established that a universal entangler exists if, and only if, $\min(m,n) \geq 3$ and $(m,n) \neq (3,3)$. Furthermore, how explicitly to construct universal entanglers to operate over an arbitrary bipartite $C^m \otimes C^n$ system is a question that remains open. An explicit example of perfect entangler for $C^3 \otimes C^4$ was given in [17] (which will be discussed in detail later), and in [18] two universal entanglers class applicable to bosonic/fermionic systems were provided.

Although do not exist an analytical formula to determine if a given unitary matrix is a universal entangler or not, there are some hints that can be followed to discard bad candidates. For example:

**Lemma 1.** *If a $n \times n$ unitary matrix $U$ is an universal entangler then all of its n columns are non-separable vectors.*

*Proof*: Let $|u_i\rangle$, a separable state, be the $i$-th column of the $n \times n$ unitary matrix $U$, and $|v_i\rangle$ is the $i$-th state of the canonical basis and, hence, it is also separable. Then $U|v_i\rangle = |u_i\rangle$, that is separable, therefore, $U$ is not a universal entangler.

A direct consequence from Lemma is that controlled gates cannot be universal entanglers. From the best of our knowledge, the only example of universal entangler in $C^3 \otimes C^4$ was proposed in [17] and corroborated in [19, 18], it is the quantum gate

$$U_H = \begin{bmatrix}
+ & - & - & - & - & - & - & - & - & - & - & - \\
+ & + & - & + & - & - & - & + & + & + & - & + \\
+ & + & + & - & + & - & - & - & + & + & + & - \\
+ & - & + & + & - & + & - & - & - & + & + & + \\
+ & + & - & + & + & - & + & - & - & - & + & + \\
+ & + & + & - & + & + & - & + & - & - & - & + \\
+ & + & + & + & - & + & + & - & + & - & - & - \\
+ & - & + & + & + & - & + & + & - & + & - & - \\
+ & - & - & + & + & + & - & + & + & - & + & - \\
+ & - & - & - & + & + & + & - & + & + & - & + \\
+ & + & - & - & - & + & + & + & - & + & + & - \\
+ & - & + & - & - & - & + & + & + & - & + & +
\end{bmatrix} \quad (9)$$

where the symbols + and − means, respectively, $12^{-1/2}$ and $-12^{-1/2}$. However, from Lemma one can see that $U_H$ cannot be a universal entangler. In fact it is easy to check that

$$U_H\left(|0\rangle_3 \otimes |0\rangle_4\right) = F_3|0\rangle_3 \otimes F_4|0\rangle_4. \quad (10)$$

In (10), $F_d$ is the single-qudit Hadamard gate in $C^d$ ($d = 3,4$), $|0\rangle_3$ is the first state of the canonical basis of a qutrit $\{|0\rangle_3, |1\rangle_3, |2\rangle_3\}$, and $|0\rangle_4$ is the first state of the canonical basis of a qudit ($d = 4$) $\{|0\rangle_4, |1\rangle_4, |2\rangle_4, |3\rangle_4\}$.

## 4. Candidates for Universal Entanglers in $C^3 \otimes C^4$ and $C^4 \otimes C^4$

To verify whether or not a given quantum gate is a universal entangler is an intractable problem, this arises from the fact that solving a system of polynomial equations is, in general, NP-hard. However, in practice, to find a counterexample that proves that a given gate is not a universal entangler is more feasible. In this work we used a Differential Evolution algorithm to enhance the ability to find counterexamples that invalidate universal entanglers candidates. Here we considered only the spaces $C^3 \otimes C^4$ and $C^4 \otimes C^4$. According to (3)-(4), a general state in $C^3 \otimes C^4$, $\alpha_1|00\rangle + \alpha_2|01\rangle + \alpha_3|02\rangle + \alpha_4|03\rangle + \ldots + \alpha_{12}|23\rangle$ is separable if

$$\sum_{i=1}^{17}|\zeta_i|=0 \tag{11}$$

where

$$\begin{aligned}
&\zeta_1=\alpha_1^*\alpha_6-\alpha_2^*\alpha_5;\ \zeta_2=\alpha_1^*\alpha_7-\alpha_3^*\alpha_5;\ \zeta_3=\alpha_1^*\alpha_8-\alpha_4^*\alpha_5\\
&\zeta_4=\alpha_1^*\alpha_{10}-\alpha_2^*\alpha_9;\ \zeta_5=\alpha_1^*\alpha_{11}-\alpha_3^*\alpha_9;\ \zeta_6=\alpha_1^*\alpha_{12}-\alpha_4^*\alpha_9\\
&\zeta_7=\alpha_2^*\alpha_7-\alpha_3^*\alpha_6;\ \zeta_8=\alpha_2^*\alpha_8-\alpha_4^*\alpha_6;\ \zeta_9=\alpha_2^*\alpha_{11}-\alpha_3^*\alpha_{10}\\
&\zeta_{10}=\alpha_2^*\alpha_{12}-\alpha_4^*\alpha_{10};\ \zeta_{11}=\alpha_3^*\alpha_8-\alpha_4^*\alpha_7;\ \zeta_{12}=\alpha_3^*\alpha_{12}-\alpha_4^*\alpha_{11}\\
&\zeta_{13}=\alpha_5^*\alpha_{10}-\alpha_6^*\alpha_9;\ \zeta_{13}=\alpha_5^*\alpha_{11}-\alpha_7^*\alpha_9;\ \zeta_{14}=\alpha_5^*\alpha_{12}-\alpha_8^*\alpha_9\\
&\zeta_{15}=\alpha_6^*\alpha_{11}-\alpha_7^*\alpha_{10};\ \zeta_{16}=\alpha_6^*\alpha_{12}-\alpha_8^*\alpha_{10};\ \zeta_{17}=\alpha_7^*\alpha_{12}-\alpha_8^*\alpha_{11}.
\end{aligned} \tag{12}$$

On the other hand, according to (3)-(4), a general state in $C^4\otimes C^4$, $\alpha_1|00\rangle+\alpha_2|01\rangle+\alpha_3|02\rangle+\alpha_4|03\rangle+\ldots+\alpha_{16}|33\rangle$ is separable if

$$\sum_{i=1}^{36}|\xi_i|=0 \tag{13}$$

where

$$\begin{aligned}
&\xi_1=\alpha_1^*\alpha_6-\alpha_2^*\alpha_5;\xi_2=\alpha_1^*\alpha_7-\alpha_3^*\alpha_5;\xi_3=\alpha_1^*\alpha_8-\alpha_4^*\alpha_5;\\
&\xi_4=\alpha_1^*\alpha_{10}-\alpha_2^*\alpha_9;\xi_5=\alpha_1^*\alpha_{11}-\alpha_3^*\alpha_9;\xi_6=\alpha_1^*\alpha_{12}-\alpha_4^*\alpha_9;\\
&\xi_7=\alpha_1^*\alpha_{14}-\alpha_2^*\alpha_{13};\xi_8=\alpha_1^*\alpha_{15}-\alpha_3^*\alpha_{13};\xi_9=\alpha_1^*\alpha_{16}-\alpha_4^*\alpha_{13};\\
&\xi_{10}=\alpha_2^*\alpha_7-\alpha_3^*\alpha_6;\xi_{11}=\alpha_2^*\alpha_8-\alpha_4^*\alpha_6;\xi_{12}=\alpha_2^*\alpha_{11}-\alpha_3^*\alpha_{10};\\
&\xi_{13}=\alpha_2^*\alpha_{12}-\alpha_4^*\alpha_{10};\xi_{14}=\alpha_2^*\alpha_{15}-\alpha_3^*\alpha_{14};\xi_{15}=\alpha_2^*\alpha_{16}-\alpha_4^*\alpha_{14};\\
&\xi_{16}=\alpha_3^*\alpha_8-\alpha_4^*\alpha_7;\xi_{17}=\alpha_3^*\alpha_{12}-\alpha_4^*\alpha_{11};\xi_{18}=\alpha_3^*\alpha_{16}-\alpha_4^*\alpha_{15};\\
&\xi_{19}=\alpha_5^*\alpha_{10}-\alpha_6^*\alpha_9;\xi_{20}=\alpha_5^*\alpha_{11}-\alpha_7^*\alpha_9;\xi_{21}=\alpha_5^*\alpha_{12}-\alpha_8^*\alpha_9;\\
&\xi_{22}=\alpha_5^*\alpha_{14}-\alpha_6^*\alpha_{13};\xi_{23}=\alpha_5^*\alpha_{15}-\alpha_7^*\alpha_{13};\xi_{24}=\alpha_5^*\alpha_{16}-\alpha_8^*\alpha_{13};\\
&\xi_{25}=\alpha_6^*\alpha_{11}-\alpha_7^*\alpha_{10};\xi_{26}=\alpha_6^*\alpha_{12}-\alpha_8^*\alpha_{10};\xi_{27}=\alpha_6^*\alpha_{15}-\alpha_7^*\alpha_{14};\\
&\xi_{28}=\alpha_6^*\alpha_{16}-\alpha_8^*\alpha_{14};\xi_{29}=\alpha_7^*\alpha_{12}-\alpha_8^*\alpha_{11};\xi_{30}=\alpha_7^*\alpha_{16}-\alpha_8^*\alpha_{15};\\
&\xi_{31}=\alpha_9^*\alpha_{14}-\alpha_{10}^*\alpha_{13};\xi_{32}=\alpha_9^*\alpha_{15}-\alpha_{11}^*\alpha_{13};\xi_{33}=\alpha_9^*\alpha_{16}-\alpha_{12}^*\alpha_{13};\\
&\xi_{34}=\alpha_{10}^*\alpha_{15}-\alpha_{11}^*\alpha_{14};\xi_{35}=\alpha_{10}^*\alpha_{16}-\alpha_{12}^*\alpha_{14};\xi_{36}=\alpha_{11}^*\alpha_{16}-\alpha_{12}^*\alpha_{15};
\end{aligned} \tag{14}$$

Given a two-qudit quantum gate in $C^3 \otimes C^4$, our algorithm searches for quantum states that minimize the left side of (11). Similarly, given a two-qudit quantum gate in $C^4 \otimes C^4$, our algorithm searches for quantum states that minimize the left side of (13). For example, our algorithm took only few seconds to find out that $U_H$ in (9) is not a universal entangler.

Although for high dimensions, as it was pointed in [17], a random unitary is almost surely a universal entangler, to build it from more familiar gates can be useful. Thus, our main goal in this work is to find good universal entanglers candidates using known gates. Here we try $F_{12}$, $X_{12}$, $Y_{12}$ and $Z_{12}$ operating over the product states $|\psi_A\rangle_3 \otimes |\psi_B\rangle_4$ and $F_{16}$, $X_{16}$, $Y_{16}$ and $Z_{16}$ operating over the product states $|\varphi_A\rangle_4 \otimes |\varphi_B\rangle_4$. The definition of these quantum gates are

$$X_d |k\rangle = |(k+1) \bmod d\rangle \tag{15}$$

$$Z_d |k\rangle = e^{i2\pi k/d} |k\rangle \tag{16}$$

$$Y_d = i X_d Z_d \tag{17}$$

$$F_d |k\rangle = \frac{1}{\sqrt{d}} \sum_{l=0}^{d-1} e^{(i2\pi kl/d)} |l\rangle. \tag{18}$$

The elements of their matrix representations are given by

$$(X_d)_{mn} = \begin{cases} 1 & \text{if } m = (n+1) \bmod d \\ 0 & \text{otherwise} \end{cases} \tag{19}$$

$$(Z_d)_{mn} = \begin{cases} e^{(2\pi i/d)^{m-1}} & \text{se } m = n \\ 0 & \text{se } m \neq n \end{cases} \tag{20}$$

$$(F_d)_{mn} = e^{(2\pi i/d)^{mn}} / \sqrt{d}. \tag{21}$$

Although all these gates are non-separable in $C^3 \otimes C^4$ (for $d = 12$) and $C^4 \otimes C^4$ (for $d = 16$), they are not universal entanglers. Our algorithm takes only few seconds to find counterexamples. The same happens to $(F_d)^{1/2}$, $(X_d)^{1/2}$ and $(Z_d)^{1/2}$. On the other hand, good universal entanglers candidates are

$$U_{E1} = \sqrt{Y_{12}} \tag{22}$$

$$U_{E2} = \sqrt{Y_{16}} \tag{23}$$

$$U_{E3} = \sqrt{X_{12}}^\dagger \cdot F_{12} \cdot \sqrt{X_{12}} \tag{24}$$

$$U_{E4} = \sqrt{X_{16}}^\dagger \cdot F_{16} \cdot \sqrt{X_{16}}. \tag{25}$$

After fifteen days looking for counterexamples for $U_{E1}$, $U_{E2}$, $U_{E3}$ and $U_{E4}$, our algorithm was not able to find anyone. The minimal entanglement value created by them can be seen in Table 1.

Table 1: Minimum entanglement generated by gates $U_{E1}$, $U_{E2}$, $U_{E3}$ and $U_{E4}$ after several days running the algorithm based on Differential Evolution.

| Quantum Gate | Minimum Entanglement |
|---|---|
| $U_{E1}$ | 0.000016199687 |
| $U_{E2}$ | 0.000057325250 |
| $U_{E3}$ | 0.005023555953 |
| $U_{E4}$ | 0.000130825518 |

Hence, our simulation results suggest the Conjecture .

**Conjecture 1.** *The gates $\sqrt{Y_{12}}$ and $\sqrt{X_{12}}^{\dagger} \cdot F_{12} \cdot \sqrt{X_{12}}$ are universal entanglers in $C^3 \otimes C^4$ while $\sqrt{Y_{16}}$ and $\sqrt{X_{16}}^{\dagger} \cdot F_{16} \cdot \sqrt{X_{16}}$ are universal entanglers in $C^4 \otimes C^4$.*

At last, we calculated the entanglement generated by the gates $U_H$, $U_{E1}$ and $U_{E3}$ when they are applied to a hundred thousand quantum states randomly chosen [20]. The estimated distributions of the entanglement values obtained are shown in Fig. 1.

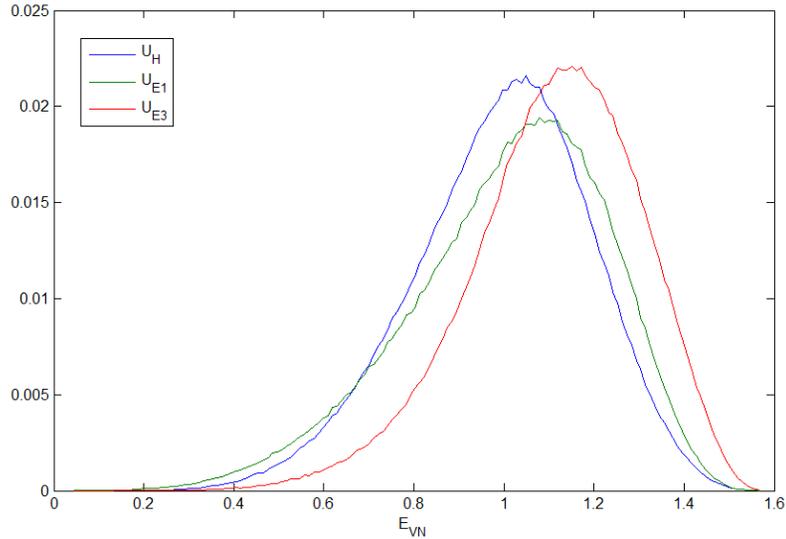

Figure 1: Distribution of the entanglement generated by gates $U_H$, $U_{E1}$ and $U_{E3}$ when they are applied to a hundred thousand quantum states randomly chosen.

The mean values of the entanglements are 0.9925, 1.0062 and 1.1096 for $U_H$, $U_{E1}$ and $U_{E3}$, respectively.

## 5. Conclusions

Using the separability conditions provided by (3)-(4) as fitness function, we constructed an algorithm based on Differential Evolution that tests if a particular quantum gate is a universal entangler. Obviously one could use an entanglement measure to test the separability, however, this requires the calculation of eigenvalues and their logarithms. Evaluating the separability using only the coefficients of the states requires a lower computational effort. The proposed algorithm showed that a candidate to universal entangler in $C^3 \otimes C^4$ found in the literature is a not a universal entangler, what can also de observed analytically by Lemma 1, and it provided two good candidates in $C^3 \otimes C^4$ ($d = 12$) and $C^4 \otimes C^4$ ($d = 16$): $\sqrt{Y_d}$ and $\sqrt{X_d}^\dagger \cdot F_d \cdot \sqrt{X_d}$.

**ACKNOWLEDGMENTS**

This work was supported by the Brazilian agency CNPq Grant no. 303514/2008-6. Also, this work was performed as part of the Brazilian National Institute of Science and Technology for Quantum Information.